# T-S3RA: Traffic-Aware Scheduling for Secure Slicing and Resource Allocation in SDN/NFV-Enabled 5G Networks

Ali J. Ramadhan[#1]

[1]Department of Computer Techniques Engineering, College of Technical Engineering, University of AlKafeel
Najaf 31001, Iraq

[1]ali.j.r@alkafeel.edu.iq

**ABSTRACT**

*Network slicing and resource allocation play pivotal roles in software-defined network (SDN)/network function virtualization (NFV)-assisted 5G networks. In 5G communications, the traffic rate is high, necessitating high data rates and low latency. Deep learning is a potential solution for overcoming these constraints. Secure slicing avoids resource wastage; however, DDoS attackers can exploit the sliced network. Therefore, we focused on secure slicing with resource allocation under massive network traffic. Traffic-aware scheduling is proposed for secure slicing and resource allocation over SDN/NFV-enabled 5G networks. In this approach (T-S3RA), user devices are authenticated using Boolean logic with a password-based key derivation function. The traffic is scheduled in 5G access points, and secure network slicing and resource allocation are implemented using deep learning models such as SliceNet and HopFieldNet, respectively. To predict DDoS attackers, we computed the Renyi entropy for packet classification. Experiments were conducted using a network simulator with 250 nodes in the network topology. Performance was evaluated using metrics such as throughput, latency, packet transmission ratio, packet loss ratio, slice capacity, bandwidth consumption, and slice acceptance ratio. T-S3RA was implemented in three 5G use cases with different requirements, including massive machine-type communication, ultra-reliable low-latency communication, and enhanced mobile broadband.*

## I. INTRODUCTION

Network slicing is an important issue that needs to be addressed for 5G networks under a single physical infrastructure. In general, network slicing is defined as selecting appropriate slices and allocating resources as per the specific service type of a user [1–4]. Currently, network equipment requires high service satisfaction rates. For example, consider a 4K ultra HD video streaming application that must meet high service requirements such as high throughput, low latency, high reliability, and high storage space. This application needs to run with large throughput and with limited delay. Network slicing and resource allocation are implemented to meet these requirements. Thus far, joint network slicing and resource allocation have been proposed as major solutions to meet users' quality of service (QoS) requirements. One of the best approaches to creating a slice is to use the service type. Data traffic consists of the type of service instance used to configure network slicing. Furthermore, the solution needs to offer a dynamic and quick response to meet the service level agreement (SLA) constraints. Most existing solutions for network slicing and resource allocation are computationally expensive and do not support mixed types of slice requests [5,6]. **Fig. 1** shows the typical diagram for network slicing in 5G. A software-defined network (SDN) provides appropriate support for network slicing and resource allocation as it has the functionality to process slice requests and perform data traffic scheduling [7]. SDNs provide several advantages in terms of network slicing and are currently being widely researched for many applications. Some of these advantages are reliable communication, urgent and long-distance communication, and solutions to optimization problems. Q-learning and Deep-Q-learning are reinforcement learning algorithms [8–10] used to identify either network traffic and/or to perform resource prediction. A heuristic algorithm is used for similar cases. However, these algorithms require more time to perform computations and are thereby highly complex.

Existing network slicing and resource allocation approaches yield poor results because of the massive arrival rate of resource requests, high data traffic in a network slice, and a large number of network slice requests [11]. Furthermore, centralized SDN controllers are affected because they act as a single point of failure and cannot handle most urgent (ultralow-latency) service requests. To tackle these issues, a multi-controller SDN environment has been established. In SDN/NFV-based network slicing, the controller makes the final decision to direct flows to appropriate slices [12]. This mechanism is vulnerable to a distributed denial of service (DDoS) attack, and therefore, authentication and bandwidth prediction are key steps that need to be followed to mitigate DDoS attacks. Hence, slice security and privacy are being researched to help manage multiple network slices in 5G infrastructures [13–15].

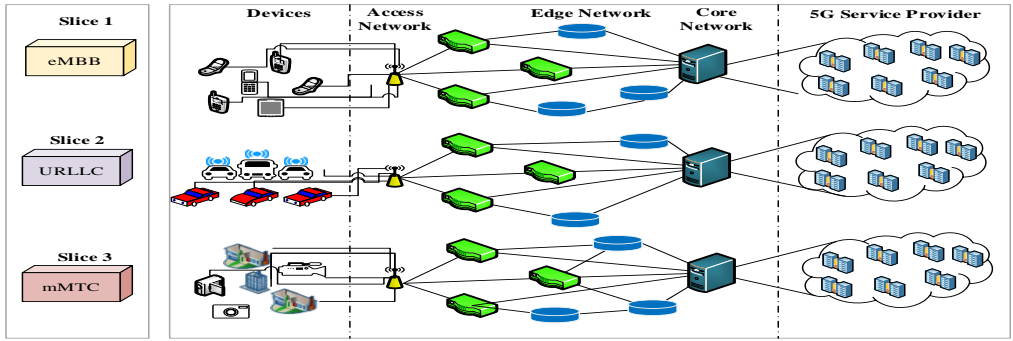

**Fig. 1: Network slice architecture**

This study focuses on handling massive heterogeneous services from diverse devices connected in 5G networks [16–22]. Three main services are currently evolving in 5G: enhanced mobile broadband (eMBB); ultra-reliable low-latency communication (URLLC); and massive machine-type communication (mMTC), as shown in Table 1.

**Table 1: 5G services**

| Service type | QoS requirements | Use cases |
| --- | --- | --- |
| eMBB | Peak throughput (20 Gb/s) | Augmented reality (AR) |
| | 1 Gb/s throughput (high mobility) | Virtual reality (VR) |
| | Minimum bandwidth (100 kbps) | |
| URLLC | Low latency (1 ms) | Smart grid control |
| | Ultra reliability ($10^{-9}$ error rate) | Industrial automation |
| | Minimum bandwidth (100 kbps) | Drone control |
| | | Robotics |
| | | Self-driving vehicles |
| mMTC | 1 million device connections | Smart cities |
| | | Asset tracking |
| | Minimum bandwidth (25 kbps) | Smart utilities |
| | | Security infrastructure |

Network slicing is enabled to support several services using heterogeneous devices. The main objective of this study is to develop specific QoS (throughput, delay, data rate, etc.) and SLA constraints (availability, reliability, etc.) for a specific service type requested across the physical network infrastructure and to assign the dedicated network slice for all incoming slice requests. This study had the following sub-objectives: (1) to design a dynamic, secure, and reliable model for network slicing and resource allocation for user equipment (UE); (2) to off-load network traffic from an overloaded switch to under-loaded switches to reduce packet loss rate; and (3) to eliminate DDoS attackers in an individual slice and prevent attackers from accessing the network.

The integration of 5G and SDN/NFV has increased the QoS and security requirements of networks. Network slicing and resource allocation in an SDN/NFV-enabled 5G network is a challenging issue because the appropriate slice and optimum set of resources need to be determined. Most studies implement network slicing and resource allocation for few heterogeneous services and consider very limited metrics. There is a need to improve network slicing and resource allocation by considering more realistic mixed types of services and improving slice capacity and UE mobility. Further, there is a need for fast network slicing and resource allocation because different service requirements have different throughputs, data rates, mobilities, and delays. A universal system would allow for improving the QoS. A centralized SDN controller is not feasible for appropriate slice selection and resource allocation. Because slice security and privacy are not considered, the participation of DDoS attackers is higher, and these attackers can send massive amounts of traffic to a specific slice. When an attacker sends a slice request, it is possible to allocate resources to the forged request. This can lead to a high packet loss rate for legitimate UE slice requests, and controllers/switches may fail at handling these large requests. Dynamic flow offloading has proved to be helpful in avoiding flow failures at switches and in eliminating packet dropping caused by static resource allocation. However, secure network slicing alone cannot satisfy the QoS requirement for the UEs because high-priority traffic needs to be served first based on service type.

We tested the proposed secure network slicing and resource allocation model for a multi-controller environment where one global controller and a set of local controllers are used. The global controller is responsible for the sliced network and for the allocation of optimum resources to obtain diverse service requirements from users. The main contributions of this paper are as follows.

To eliminate DDoS attackers from network slices and secure entire slices, we perform device authentication in 5G access points (APs), which consist of a set of virtual authority pools (VAPs). Each authority in the VAP is audited by the 5G AP and can insert, delete, and modify operations. A Boolean logic-based password-based key derivation function 2 (PBKDF2), which uses three input parameters—a secret key, a physically unclonable function (PUF), and a timestamp—is used. We perform traffic scheduling for devices using an asymmetric queue model that functions based on Bernoulli's theorem. To schedule traffic flow, we use data rate, packet delay, and packet length. Two queues are fixed in the 5G AP: a high-priority queue and a low-priority queue. The service rates of each queue are asymmetric with respect to the arrival rate. For network slicing, we use SliceNet, which is a lightweight and faster convolutional neural network (CNN) that outperforms ByteNet, WaveNet, and traditional CNNs. For slicing the network, we consider traffic type, fair SLA, international mobile subscriber identity (IMSI), slice capacity, and device mobility. In this work, we primarily focus on SLA constraints between the user and the service provider, which results in fair SLA while slicing the network. We first compute the service availability ratio (SAR), response time ratio (RTR), throughput ratio (TR), and service reliability ratio (SRR), and then the fairness (weight) for each service. Resource allocation is performed using HopFieldNet, which is a fast neural network that finds resources for each slice. To handle overloading at network slices, we propose a dynamic flow offloading scheme that can be employed at the local control plane. We use a fast-weighted bipartite graph ($F\omega BG$) based on switch service capacity, transmission rate, and loss rate; this graph maps multiple flows to the optimum switches that can improve the network reputation. In addition to device authentication, we implement packet classification using Renyi entropy. Here, the bandwidth usage is predicted for the switches. Finally, experiments are conducted using the NS3.26 simulator to evaluate the proposed scheme, which exhibits highly satisfactory performance compared with previously proposed schemes in terms of several metrics, such as throughput, latency, response time, packet transmission ratio, packet loss ratio, slice capacity, bandwidth consumption, and slice acceptance ratio.

The remainder of this article is organized as follows: Section II reviews the earlier research on network slicing and resource allocation to identify the research gap. In Section III, the main problems discovered from existing studies are highlighted. In Section IV, the proposed T-S³RA architecture is detailed with the essential algorithms. In Section V, we compare the proposed architecture with preceding research based on experimental findings. In Section VI, we present our contributions and future enhancements.

## II. RELATED WORK
### A. Secure Network Slicing

Secure slicing allows multiple users to access a single network after authentication. Each slice security is handled in an appropriate manner because bandwidth is an important resource for enriching the network QoS. For example, vehicle-to-everything (V2X) networks are applied over 4G with long-term evolution (LTE). Furthermore, user behaviors on the slice are evaluated by verifying attributes such as the presence of malware and the strength of passwords [23]. In one study [24], the network security focused on mitigating DoS attacks in an SDN. This is achieved via switch bandwidth congestion prediction. A comprehensive judgment score is computed for each switch that represents attack severity. Using trust values, the priority manager can manage multiple buffer queues based on the different priorities of the users. To schedule flow requests from users, a weighted round-robin algorithm is employed. Although the compromises by DoS attackers are predicted, authenticating the users will help achieve attack prevention. This step can easily eliminate attackers before the controller is overloaded.

The VIKOR multicriteria decision-making (MCDM) approach was proposed in a previous study [25] for network slicing in a 5G environment. This approach finds node significance (resource and topology attributes) and ranks the nodes accordingly. Among the slice nodes, a candidate physical path is determined to maximize the slice acceptance ratio. However, VIKOR has one big drawback: the slice acceptance ratio, which must be large for high-priority traffic, is marginal.

Secure keying is established when the slice is accessed by third-party application services [26]. This method guarantees consent from the monitored devices and security properties for the keying scheme to demonstrate the security feature under a 5G services establishment. Initially, cryptographic keys are generated from the key distribution server (KDC). For key generation, the ELGamal Cryptosystem, in which two sets of keys—public and private keys—are generated, has been proposed. Here, the resources are constrained for key generation.

### B. Network Slicing and Resource Allocation

Optimum workload allocation for distributed 5G-based SDN/NFV networkshas been proposed [27]. An end-to-end network slicing architecture is designed to support many services such as eMBB and URLLC. In addition, the network operating cost can be reduced by slicing the requests from clients in the combined environment (SDN, NFV, and edge computing). In a previous study [28], the authors performed network slicing in a large-scale Internet of Things (IoT) environment (long range wide area network). Three slices of network are segregated: the urgency and reliability-aware slice, reliability-aware slice, and best-effort slice. First, cooperative slicing is implemented using one-to-many matching (determined by the number of IoT devices assigned to virtual slices). Then, resources are allocated for each slice (inter-slice resource allocation) using a one-to-one matching game. The coalitional multigame theory consumes high processing time and leads to a high computational complexity.

Packet-based data traffic scheduling has been proposed and is considered to improve resource allocation and sharing in 5G slice networks [29]. Two modes of operations are used: static sharing resource (SSR) and dynamic sharing resource (DSR). For resource allocation, the assigned capacity weight is determined for each slice and allocated accordingly. In the final analysis, the fairness for resource distribution per slice is computed. To operate massive types of slices, such as popular, heavy, and sensitive slices, a global network controller is required. Radio resource management (RRM) for multiple slice management has been investigated in a previous study [30]. In the aforementioned study, the RRM function was used to divide the radio resources and then allocate them. An interaction between the eNB and the SD-RAN controller proved slicing and allocation provisioning. A single controller cannot meet the QoS requirements for real-time traffic, and it does not suit complex application scenarios (AR/VR and Drone Control). Researchers have also proposed a slice management scheme that assigns resources based on priority [31]. High-priority slice requests are forwarded to the shortest paths and lower-priority slices are transmitted to other paths. Experiments are conducted using 200 nodes; the nodes are deployed using a grid network topology. In the final analysis, the average throughput slices up to 6, 13, and 7% and the delays of slices are minimized by up to 11–14%. Flow (high priority) forwarding in the data plane uses the shortest path. However, the shortest path is not available for all cases. Therefore, when the flow does not match with a flow table, the corresponding flow must be installed in the controller. This leads to a bandwidth scarcity problem for users under static resource allocation.

In a previous study [7], a network slice embedding model was presented; this model operates based on reliability. The model simultaneously increases the number of slice requests and reduces the failure rate of the network slices. The study uses the Lyapunov optimization model that allocates resources and ensures queue stability. It effectively improves network throughput and guarantees network stability and reliability. However, it is difficult to obtain an abundance of network slices and the interoperability between the 5G network and SDN/NFV is low. Service function chaining has been used for network slicing [32]. Each slice contains a set of service function chains, and it deals with any type of traffic per slice. Then, a greedy-based heuristic algorithm is designed for examining the trade-offs between slicing and execution runtime. Finally, an optimization model is provided to achieve the required bandwidth and delay. The mobility of network slices is not considered, and it decreases the QoS and QoE. A network slicing resource management (NSRM) scheme has been developed to allocate resources for each slice in a network [33]. This study considers an LTE network for different slice assignments and fair bandwidth distribution among slices. For each slice, the controller is deployed, and the LTE slice controller handles all slices. Through the LTE slice controller, the radio network resources are allocated via a virtual eNodeB. The dynamic provisioning of slice requests is very difficult. For example, the Industry 4.0 application requires large slice request handling.

Service-aware multiresource allocation has been performed for software-defined cellular networks in a previous study [34]. In the aforementioned study, joint network allocation and scheduling of available network resources to determine the network slices is demonstrated. Network resources are provided according to the SLA constraints and priorities. For each priority, resources (throughput, latency, storage, and reliability) are computed using the analytic hierarchy process (AHP). The experiments are tested for vertical industries and subscribers in a cost-efficient manner. The network slice capacity is a key element for determining the required resources. In addition, the AHP cannot support multiple traffic flow resource allocation concurrently. A dynamic end-to-end slicing approach has been used to test the vehicular ad hoc network environment with support for 5G communications [35]; two types of slice services are tested: Video and Web. In a single physical network infrastructure, resources are allocated for network slices. In both the data and control planes, virtualized network functionality customization is enabled to support the handling of different services from numerous users. This end-to-end approach does not provide high throughput for limited latency service types.

In a previous study [36], two types of services were examined, namely, eMBB and V2X. Two approaches— reinforcement learning and heuristic algorithm—are used for network slicing and resource allocation for different slices. The simulation results reveal that it reduces latency by 0.18 s (by 0.26 s for fixed slicing ratio). This approach cannot support multiple traffic classes in each slice. In addition, a global controller is required to reduce latency further for each service type. A quick and optimal response approach has been proposed for resource slicing in heterogeneous cellular networks [37]. First, this approach captures the real-time arrival of slice requests using a semi-Markov decision process (deep double dueling) that predicts service time and resources. Several experiments were conducted to demonstrate the performance of the proposed deep dueling approach for resource slicing. Dynamic and large network slices pose several challenges in the control and data plane. To address these challenges, load balancing among multiple network service chains has been presented. A new concept, i.e., point of presence, is used. It solves the scalability problem and accepts only a limited number of slice users [38–40].

### III. PROBLEM DEFINITION

In SDN/NFV-based 5G networks, network slicing and resource allocation are implemented in accordance with the service requirements of users or devices [41–43]. A multiclass queueing and traffic analysis model has been developed to handle massive service requests from the user. Low-complexity traffic predictors using a soft gated recurrent unit (GRU) are employed, and resources are allocated using deep neural networks (DNNs).

Furthermore, a multistage analysis is performed for three different slices (eMBB, URLLC, and mMTC); these slices are used to perform queueing based on M/M/n/K. The main limitations of the aforementioned studies are as follows: 1) The studies are based on the load; however, because the response time for processing high-priority class packets is high, it can cause non-real-time traffic to be scheduled before real-time traffic. 2) To schedule particular slices, a first-come-first-serve (FCFS) protocol is followed, which increases the response time further and leads to poor QoS for arrived requests. 3) In an SDN, a single point of failure occurs when requests from several users cannot be handled by a single controller. 4) For more realistic and QoS-constrained traffic, running both DNN and GRU increases the response time. In addition, the DNN consumes a large amount of energy and has a long training time. 5) The FCFS runs a random seek pattern because it does not reorder requests by slice to minimize service delay. Furthermore, the queuing theory does not use fair level SLA constraints. Since the service must be delivered with 99.99% availability, there is a need to ensure <0.01% service timeouts and that 99.99% of the services are completed within the resources. Furthermore, when using FCFS for scheduling, resources are not properly distributed.

A dynamic flow migration has been proposed for embedded services under SDN/NFV-assisted 5G networks, which decreases the dynamic traffic load per slice [44]. This issue has been addressed using a heuristic algorithm. The drawbacks of this approach are follows: (1) Adaptive flow migration is required because a Poisson traffic model is limited in its performance. (2) The routing path for flow migration requires an optimum solution and delay sensitive traffic cannot be executed in former methods. (3) When an unexpected flow arrives at the controller, the heuristic algorithm cannot provide an optimal solution. These aforementioned problems are resolved in the proposed research using network slicing, resource allocation, dynamic offloading, security, and packet classification.

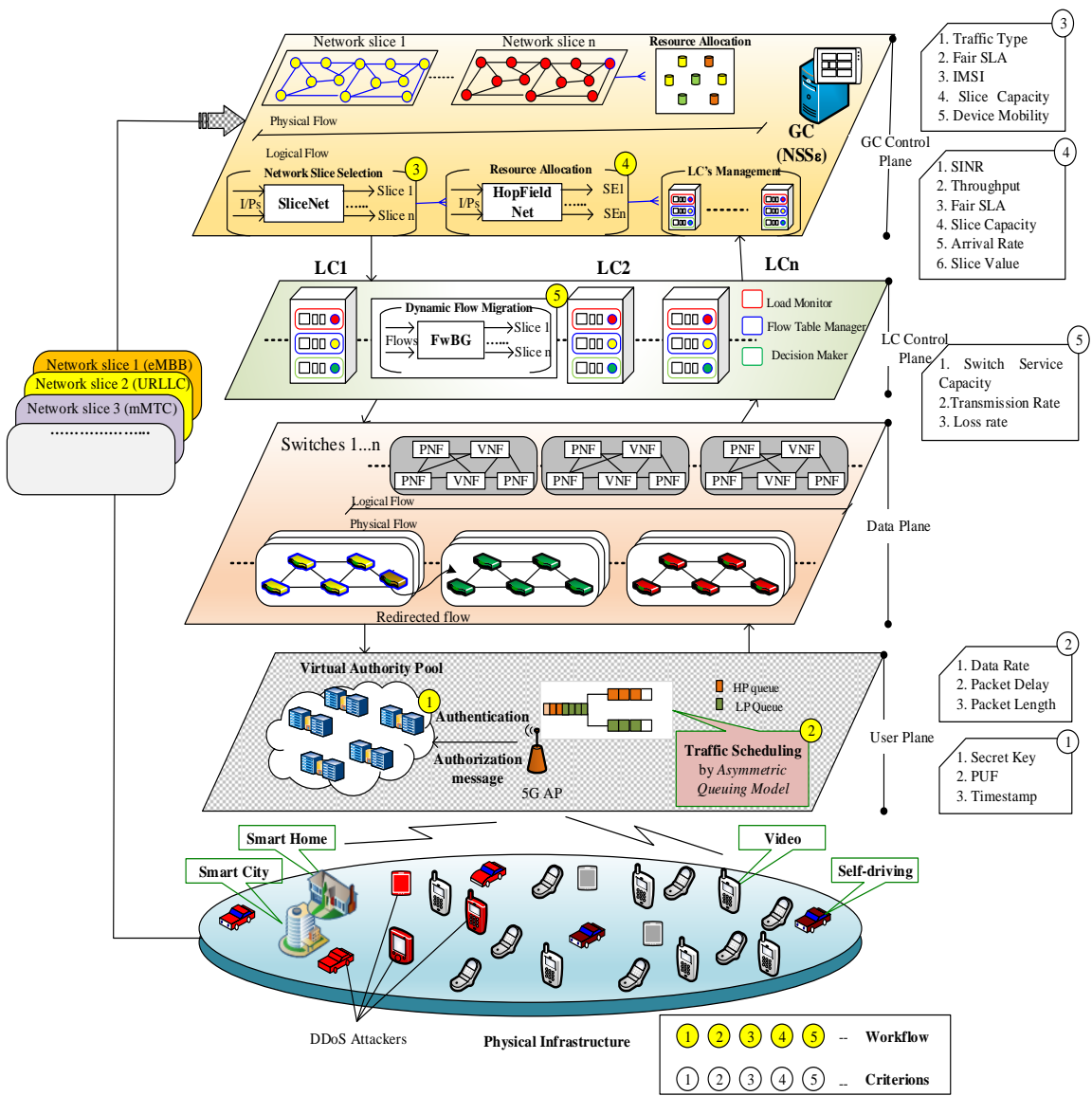

**Fig. 2: System architecture**

## IV. SYSTEM MODEL

The proposed T-S³RA in an SDN/NFV-enabled 5G network was designed using the processing capabilities of mobile device authentication, traffic scheduling, network slicing, and resource allocation and dynamic offloading.

### A. Network Overview

We designed our T-S³RA model for secure network slicing and resource allocation, which comprise the following planes: user plane, data plane, local control plane, and global control plane. The proposed T-S³RA model contains a set of entities, including devices ($d_1 \ldots d_n$), 5G APs ($AP_1 \ldots AP_n$), VAP ($VA_1 \ldots VA_n$), switches ($PS_1 \ldots PS_n$ and $VS_1 \ldots VS_n$), controllers (GC), and a set of LCs ($LC_1 \ldots LC_n$). The devices submit secure credentials to the 5G AP. In general, the components at the data plane and the control plane are limited in terms of resources. Furthermore, these resources must be used to send and receive responses from users and for action processing. The VAP is implemented for verifying device credentials, and it facilitates the reduction in complexity issues. When the VA is not required, it is removed. Thus, the single-controller-failure problem is resolved because of the use of multicontrollers and because both physical and virtual switches are used. This also helps solve the overload problem.

In network slicing, massive traffic induces more delay and packet losses. Hence, traffic is scheduled, slicing is performed, and resources are allocated using deep learning approaches. To avoid imbalance issues, dynamic offloading actions are considered. These actions are stored in the data plane by various credentials. To consider the resource wastage problem, DDoS attackers that arrive at the switches are detected using entropy calculations. The proposed T-S³RA architecture is depicted in **Fig. 2**. The primary network entities are defined as follows. (1) Devices: The set of all IoT devices that access the network via the 5G communications network are called the devices. These devices are heterogeneous in nature and dynamic in movement. Its connection to the 5G AP needs high coverage. All devices are not authorized, and there is also a chance of unauthorized user participation. (2) 5G AP: It is equipped with high communication coverage with a high data rate, high throughput, and low latency. These highly configured 5G APs are deployed in the network along with their trusted parties, which validate devices for effective authentication. (3) VAP: It is not a single entity but a pool that comprises a number of virtual authorities. It is often referred to as a special entity that balances the process of authentication. The process of VA creation and deletion is handled by the 5G AP. (4) Switches: In the data plane, switches are employed as a common that functions by matching the incoming flow with the flow table and then performs actions accordingly. These switches are virtual and physical to balance the load in the case where packets arrive into the data plane. (5) Controllers: Distributed controllers are deployed in the control plane for network slicing and resource allocation. To avoid the problem of a single point of failure, multiple controllers are used in this study.

### B. Device Authentication

To eliminate DDoS attackers from network slices and preserve the privacy for the entire network, we first implement device authentication in 5G AP, which consists of a set of VAs. Each authority in the VAP is audited by the 5G AP and has a charge to insert, delete, and modify operations. The PBKDF2, which uses three input parameters—secret key $\varsigma_r$, PUF $puf$, and Timestamp $\dagger$—is employed for authentication. If the three parameters are valid, then the request is accepted. Otherwise, it is not accepted and is terminated. These operations are performed in the Boolean logic function.

**Fig. 3** depicts the Boolean logic-based authentication for devices. The procedure for PBKDF2 is as follows: PBKDF2 is a key-derivation function created by RSA Labs. It overwhelms brute face attacks caused by weak user passwords. A PBKDF2 is derived using the following parameters: a pseudorandom function, $PRF$; an iteration count, $i_\varsigma$; a password, $pw_d$; a salt, $\mathcal{Z}$; a chosen output key length, $OK_l$; and an output-derived secret key, $\varsigma_r$.

The PBKDF2 derives a $\varsigma_r$ of arbitrary length. In particular, the PBKDF2 generates as many blocks $t_i$ as required to cover the output secret key length. Each block, $t_i$ is calculated for $PRF$ iteration using the count, $i_\varsigma$. For a large secret key length, we can add any number of iterations. Here, $i_\varsigma = 1,000$ for the generation of any critical keys. In PBKDF2, the inputs are the user password, $pw_d$1 salt values, $\mathcal{Z}$; iteration count, $i_\varsigma$; $puf$; timestamp, $\dagger$, and chosen output key length, $OK_l$. The output is a secret key, $\varsigma_r$.

$$\varsigma_r = PBKDF2\ (pw_d, OK_l, \mathcal{Z}, i_\varsigma), \qquad (1)$$

where $\varsigma_r$ is a concatenation of all security credentials. Here, PUF-based authentication is handled via two steps of processes—enrollment and verification. In the enrollment, the VA stores all challenge and response pairs of the device. It is verified whenever a device enters the network. For the verification, the verifier receives the device ID and determines the *Challenges Response Pair*, which is random. Then, the challenge of this entity is forwarded. For the issued challenge, the corresponding response is computed. The verifier checks if the response in the database and the generated response are valid; if they are, $\varsigma_r$ is generated.

As shown above, the secret key is generated to verify the authorization of all devices. Here, the Boolean logic operator is denoted as $o$, and the expression is

$$o = \overline{(\dagger \cdot puf) \cdot (\dagger + puf)} \cdot \varsigma_r. \qquad (2)$$

Based on the Boolean values, it stays connected to the network and serving. If the authentication is successful at VA, the status report is updated to the 5G AP, and the next action is performed. Otherwise, the authenticated request is terminated, and the packets are dropped at the 5G AP. Based on the response of the 5G AP, the packets are forwarded.

*C. Traffic Scheduling*

To avoid congestion at the SDN controller, we perform traffic scheduling in the 5G AP. In this step, traffic flows from devices are classified and scheduled using an asymmetric queue model that operates based on Bernoulli's theorem [45, 46]. To schedule traffic flow, we use three parameters: data rate $\tau_\gamma$, packet delay $\rho_d$, and packet length $\rho_l$. Two queues are fixed in the 5G AP: the high-priority queue (HP) and the low-priority (LP) queue. The service rate of each queue is asymmetric with respect to the arrival rate. We assume that the HP queue is an inelastic flow and that the LP consists of elastic flows. The traffic rates of the HP and LP queues are $\mu_1$ and $\mu_2$, respectively. Hence, the total queueing service rate is computed as follows:

$$\mu_1 + \mu_2 = 1. \quad (3)$$

To obtain a zero-packet loss rate, we focus on adaptive queue running between the two queues. Hence, the service rate of the HP is $\delta = 0.75$; when this value is reached and exceeded, the queue is running. Otherwise, the LP queue is running until the value of HP is 0.

The Bernoulli theorem is related to fluid dynamics, i.e., it considers the pressure, density, and flow speed as parameters of a flow. Two types of flows are considered: compressible and incompressible flows. The following three properties from Bernoulli's theorem are important and are considered in this study: (1) the flow must be steady at any point and cannot change with time; (2) the flow must be incompressible even though pressure varies, and the density must remain constant along the streamline; and (3) the friction caused by viscous forces must be negligible.

In this paper, we consider a discrete time system to schedule services that arrive dynamically for slicing requests. All arriving requests are diverse with respect to the type of service and the slots. Here, the main aim is to respond to the service requests and deliver the response within the allocated slots. Hence, we focus on traffic scheduling. All devices demand $\omega$ amount of service in slots $T$. Let $n(T)$ be the number of service requests in the HP queue at time $T$. We then define a random variable $\gamma(T)$ that represents the queue current state as

$$\gamma(T) = \begin{cases} 0 & \text{Queue is idle} \\ i & \text{Queue is busy with n services} \\ l+1 & \text{Queue is vacant} \end{cases} \quad (4)$$

When the queue size becomes $\gamma (\gamma \geq 1)$, it starts the service scheduling of the devices. Otherwise, it remains idle. Initially, it is assumed that the response time of the service from the queue in each step is distributed. After service time $\gamma (\gamma \geq 1)$, the queue starts instantly and sends $l$ successive steps to each device. This implies that the first service request from the device is followed by the second step of service. The second step of service is followed by the third and so on, up to $l$ steps of service. Soon, all service requests at step $l$ are computed for $n$ devices. At this point, the queue probability rate is $1 - P$, and it then goes to a vacant queue.

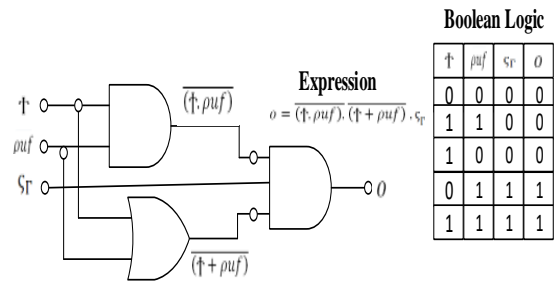

**Fig. 3: Boolean logic with PBKDF2**

| Pseudocode for Boolean logic based PBKDF2 Authentication |
|---|
| Input: Device Credentials |
| Output: Authentication (Accepted / Not) |
| Begin |
| For $\forall d$ //Registration started |
| Register $\{ID, pw_d, OK_l, 2, i_c, puf, and \dagger\} \to VA$ by 5G AP |
| At 5GAP generates $\varsigma_r$ using PBKDF2 |
| At 5GAP provides $\{\varsigma_r\} \to d$ |
| End for //Registration completed |
| If $d$ needs to access network //Authentication |
| Compute $\varsigma_r$ using (1) |
| $d_i$ submits $\{ID, pw_d, OK_l, 2, i_c, puf, and \dagger\} \to VA$ |
| $VA$ confirms credentials |
| **Forward results to Boolean logic** |
| If (Credentials are VALID) |
| $d_i = 1 \ (YES) \ Acccepted$ |
| Else |
| $d_i = 0 \ (NO) \ Not \ Accepted$ |
| End if |
| Else |
| End process |
| End if |
| End |

*D. Network Slicing and Resource Allocation*

In this section, the network slicing and resource allocation functionalities are elaborated on. We tested our secure network slicing and resource allocation model in a multicontroller environment where we deployed one GC and a set of LCs. The GC is responsible for slicing the network and allocating the optimum resources for diverse service requirements from users. We first elaborate on the network slicing and then discuss the resource allocation. In the GC, we fed network slice selection entities ($NSS^\varepsilon$s) for slicing the network; we used SliceNet for network slicing. It is a lightweight and faster CNN that outperforms ByteNet, WaveNet, and traditional CNNs.

For slicing the network, we consider Service Type $s_t$, Fair SLA $f_{SLA}$, $IMSI$, Slice Capacity $S_c$ and Device mobility $d_m$. In this work, we primarily focus on SLA constraints between the user and service provider, which results in $f_{SLA}$ while slicing the network. We compute SAR, RTR, TR, and SRR and then compute the fairness (weight) for each service.

During network slicing for a device, a service request is forwarded to the controller via the 5G AP. The proposed SliceNet is expressed as a mapping from the input layer to the output layer and can be expressed as

$$y = f(s_t, f_{SLA}, IMSI, S_c, d_m \ \forall(d_i)), \quad (5)$$

where $d_i$ is the device i. The aforementioned parameters are fed into the proposed SliceNet as inputs. In SliceNet, the input encoder, I/O mixer, and decoder are the main components. In the input encoder, the conv_module is employed; in the I/O mixer, the attention module is used; and in the decoder, both the conv and attention modules are used.

A convolutional module obtains input from the devices. This is performed in three steps: ReLU activation, separable conv, and normalization of the layer. In the normalization, hidden units are computed and normalized layer-wise. **Fig. 4** shows the SliceNet architecture. Overall, the conv_module can be represented as

$$ConvStep(w, x) = LN(SepConv.(w, ReLU(X)), \quad (6)$$

where $LN$ is the layer normalization; $w$ is the learning weight; and $SepConv$ is the separable convolution. Therefore the, conv_module can be obtained by stacking four convolutional steps:

$$h1(X) = Conv\_Step(w_{h1}^{3\times1}, X), \quad (7)$$

$$h2(X) = X + Conv\_Step(w_{h2}^{3\times1}, h1(X)), \quad (8)$$

$$h3(X) = Conv\_Step_{1,1}(w_{h2}^{15\times1}, h2(X)), \quad (9)$$

$$h4(X) = X + Conv\_Step_{1,1}(w_{h2}^{15\times1}, h3(X)), \quad (10)$$

$$Conv_{Module}(X) = \begin{cases} Dropout(h4(X), 0.5) & Training \\ h4(X) & Otherwise \end{cases}, \quad (11)$$

where $h(1 \dots n)(X)$ is the number of hidden units, and 0.5 is the learning rate.

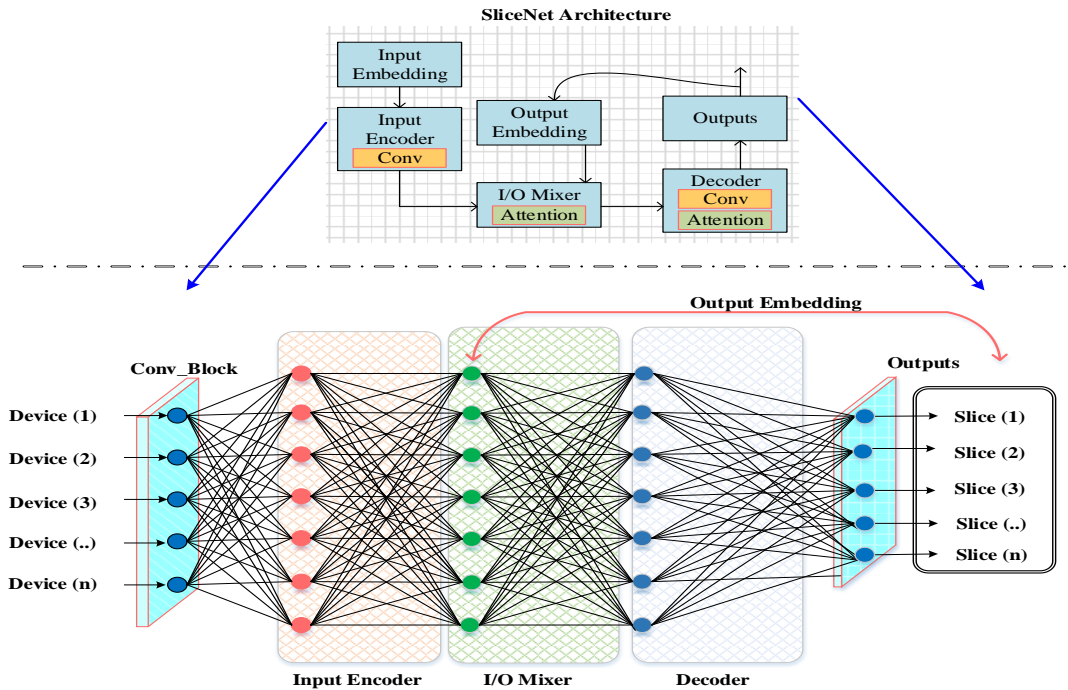

**Fig. 4: SliceNet architecture**

In the attention module, all input parameters for slices are considered as inputs, and then, the input feature vector similarities and service requirements are computed according to the service type. In this module, two convolution steps are performed by the $Softmax$ function.

$$attention\ 1(X) = ConvStep_{1,1}(w_{a1}^{5\times1}X + Time), \quad (12)$$

$$attention\ (Sr,Ta) = Attend\ (Sr,Ta,Convstep_{4,1}(w_{a1}^{5\times1}, attention-1(Ta)) \quad (13)$$

Finally, we detail the structure of the three components: input encoder, I/O mixer, and decoder. The output embedding is the concatenation of all the aforementioned components:

$$Mix_i = I/O\ (m)[I_e\ (i), O_e(o)], \quad (14)$$

$$Outputs = Decoder\ (mix). \quad (15)$$

Finally, in the output layer, we obtain $y = (y(1), y(2), \ldots y(n))$, where $y(i) = 0,1$ is defined as the slice selection indicator. Hence, $y$ represents three types of services, which are eMBB, URLLC, and mMTC with various resource configurations. Therefore, each service type, along with the specified network slice, is represented as $\mathbb{S}_j = 1\ldots n$. Then, resource allocation is implemented using HopFieldNet, which is a fast neural network that finds the resource for each slice. To determine the resources for each slice request, we use $SINR$, $Throughput$, $f_{SLA}$, $\mathbb{S}_c$, arrival rate $A_r$, and slice value $s_v$. Here, the resources are allocated for three different processes—communication, computation, and caching—which are denoted by $c_i$, $c_j$, and $c_k$, respectively.

HopFieldNet is a type of artificial neural network (ANN) that is composed of nodes on a single layer. The input nodes in HopFieldNet are updated synchronously based on clocktime variations. Here, the contributing nodes exist with the connectivity based on the determined weight values. HopFieldNet uses the results from network slices as input and computes the resources for the three classes of slices as eMBB, URLLC, and mMTC. The opinion loops designed in this network imitate its performance in enriching knowledge capability. This is effective in resolving complex computational issues. HopFieldNet is designed with a single layer of input nodes that are connected to other nodes as feedback connections that assist in redirecting the output to the input. Here, the number of nodes, inputs, and outputs are equal and, in this T-S³RA system, the total number of resources $R_1, R_2, R_3, \ldots, R_N$ nodes are built.

The nodes are measured to be double-threshold nodes since they are served in a content-addressable memory system. As stated by the arrival of input, it describes an equivalent weight value. The weight value of the received input is strong-minded from the individual slice service requirements that are expressed based on the weight values in the connection and also on the state of the node. The weighted sum of the nodes $U$ is

$$U_i = \sum_{j=1}^{N} we_{ij} st_j, \quad (16)$$

where $we_{ij}$ represents the connectivity weight existing between $i$ and $j$, and $st_j$ is the state of the node $j$. **Fig. 5** shows a pictorial representation of HopFieldNet. The training in HopFieldNet is controlled using the Storkey learning rule, which is applied for better error minimization. Mathematically, the Storkey learning rule is formulated as

$$we_{ij}^0 = 0 \quad \forall\ i,j, \quad (17)$$

$$we_{ij}^k = we_{ij}^{k-1} + \frac{1}{N}\xi_i^k\xi_j^k - \frac{1}{N}\xi_i^k h_{ji}^k - \frac{1}{N}h_{ij}^k\xi_j^k. \quad (18)$$

This learning rule allows the local and incremental properties to update the connectivity weight data and upturns if there is no need for information from any other before the trained pattern. In (17) and (18), $w_{ij}^{ek}$ is the weight estimated between $i$ and $j$ only after the $k^{th}$ pattern is learned, while $\xi^k$ denotes the new knowledge pattern. The local field $H_{ij}^k$ is given as

$$H_{ij}^k = \sum_{n=1, n\neq i,j}^{N} we_{in}^{k-1}\xi_n^k. \quad (19)$$

HopFieldNet is premeditated to categorize the resources based on the service requirements of the slice.

From **Fig. 5**, we show that the HopFieldNet with a single layer for classifying the available resource blocks from the slices are $\{x_1, x_2, \ldots x_i \ldots, x_N\}$ and the corresponding outputs are $\{Y_1, Y_2, y_3, \ldots Y_i \ldots, Y_N\}$. The inputs are received from all NS, which are determined as $\{R_1, R_2, R_3, \ldots, R_N\}$. The output in HopFieldNet is attained for each separate NS. **Table 2** summarizes the resource allocations for HopFieldNet.

**Table 2: HopFieldNet-based resource allocation**

| Service | Service indicator | Resources |
|---------|-------------------|-----------|
| eMBB    | (0, 0, 1)         | $S1 = c_i, c_j, c_k$ |
| URLLC   | (0, 1, 0)         | $S2 = c_i, c_j, c_k$ |
| mMTC    | (1, 1, 1)         | $S3 = c_i, c_j, c_k$ |

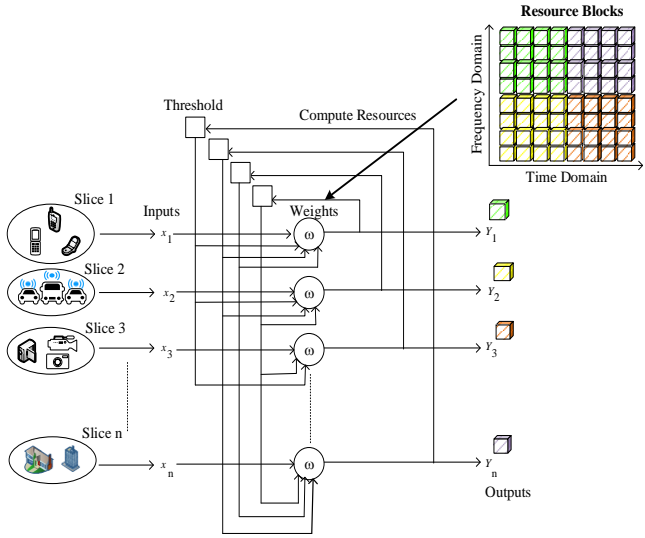

**Fig. 5: HopFieldNet**

The benefit of using HopFieldNet is its procedure for associative memory. This memory is aided to store part of the information, using which the rest of the pattern can be recollected. The recall of the previous patterns allows for the use of prior knowledge of the resource amount for each NS. The resource is categorized into three states, as summarized in **Table 2**. The states of nodes in the proposed T-S³RA are estimated as

$$St = (s_{t1} \quad s_{t2} \quad \cdots \quad s_{ti} \quad \cdots \quad s_{tN}). \quad (20)$$

The states $st$ for each node are formulated in a matrix that is trained, and the three classes, $c_i$, $c_j$, and $c_k$, are the possible states of the resource. The state of node $st_i$ is determined as follows:

$$st_i = sign(U_i - THRES_N), \quad (21)$$

where $THRES_N$ denotes the threshold, $sign(x) = 1 \forall X \geq 0$, and $sign(x) = -1 \forall X < 0$. Then, the weighted values for each node in the constructed HopFieldNet are determined in a matrix with a zero diagonal. As in this network, no node is associated with itself, all nodes should obey $w(e) = w(e)$, and the weights of node connectivity are expressed as

$$WE = \begin{pmatrix} 0 & w(e)_{12} & \cdots & w(e)_{1i} & \cdots & w(e)_{1N} \\ w(e)_{21} & 0 & ,\cdots & w(e)_{2i} & \cdots & w(e)_{2N} \\ \vdots & \vdots & \ddots & \vdots & \cdots & \vdots \\ w(e)_{i1} & w(e)_{i2} & \cdots & 0 & \cdots & w(e)_{iN} \\ \vdots & \vdots & & \cdots & \ddots & \vdots \\ w(e)_{N1} & w(e)_{N2} & \cdots & w(e)_{Ni} & \cdots & 0 \end{pmatrix} \quad (22)$$

Threshold $THRES_i$ is given for each node according to its service requirements. Therefore, it is not necessary to have all listed services in each Ns. The threshold for the nodes is given in the matrix format as

$$THRES_N = \begin{pmatrix} \theta_1 \\ \theta_2 \\ \vdots \\ \theta_i \\ \vdots \\ \theta_N \end{pmatrix}, \quad (23)$$

where $\{\theta_1, \theta_2, \ldots \ldots \theta_N\}$ are the individual threshold values for each node. Based on the presence of the slice requests in each NS, the threshold can be varied. If the user or device includes a new slice request, then the threshold is updated. After detecting the resources for the NS, the utilization of the category of the resources at the NS is identified. Then, the individual NS payment status is verified, and the exact utilization of the load by the NS is predicted.

*E. Dynamic Flow Offloading*

An insufficient bandwidth of switches can lead to a higher traffic volume of slices. To handle overloading at the network slices, we propose a dynamic flow offloading scheme at the local control plane. We use $F\omega BG$ based on the switch service capacity, transmission rate, and loss rate. The $F\omega BG$ maps multiple flows to the optimum switches that will increase network reputation. Further, the use of $F\omega BG$ helps avoid slice capacity problems. In addition, we find the compromised DDoS attackers in switches via packet classification using service requests arrival rate, request time interval, and packet size. **Table 3** lists the notations used in this paper.

**Table 3: Notations used in this paper**

| Notations | Description |
| --- | --- |
| $d = d_1, d_2, \ldots, d_N$ | n number of devices |
| $AP = AP_1 \ldots AP_n$ | $n$ number of access points |
| $VS_1 \ldots VS_n$ | $n$ number of virtual switches |
| $PS_1 \ldots PS_n$ | $n$ number of physical switches |
| $VA_1 \ldots VA_n$ | $n$ number of virtual authorities |
| $LC_1 \ldots LC_n$ | Set of local controllers |
| GC | Global controller |
| $puf$ | Physically unclonable function |

| | | | | |
|---|---|---|---|---|
| $D_{ID}$ | Device identity | $T$ | Timeslots |
| $P_{wd}$ | Password | ɣ(T) | Random variable |
| $ST$ | Session token | $LN$ | Layered normalization |
| $ς_Γ$ | Secret key | $NS$ | Network slices |
| † | Timestamp | $c_i, c_j, c_k$ | Communications, computing, and caching |
| $PRF$ | Pseudorandom function | | |
| $i_ς$ | Iteration count | | |
| $ς$ | Salt | | |
| $OK_l$ | Output key length | | |
| HP and LP | High priority and low priority | | |
| $\mu_1$ and $\mu_2$ | Service rate | | |

## V. EXPERIMENTAL RESULTS AND DISCUSSION

### A. Simulation Setup

The proposed T-S³RA model was investigated using the network simulator tool version NS3.26. This tool is capable of incorporating the network modules and technologies required for appropriately simulating a network. The network simulator was installed on a system with the Ubuntu 14.04 LTS OS, a 32-bit dual-core processor, and 2 GB of RAM. The simulation parameters used to design the testbed are listed in **Table 4**.

**Table 4: Simulation settings**

| Simulation parameter | Specification |
|---|---|
| Area size | 1000 m × 1000 m |
| Simulation tool | NS3 |
| Simulation time | 300 s |
| **Network Settings** | |
| Number of devices | 250 |
| Number of illegitimate devices | 5–10% |
| Number of VAs in VAPNumber of APs | 2 |
| Number of TTP | 1 |
| Number of switches | 1 |
| Number of local controllers | 38 (OpenvSwitch 0.8.9) |
| Number of global controllers | 3 |
| SDN controller | 1 |
| Cloud service provider | POX |
| | 1 |
| **Flow Settings** | |
| Flow type | TCP, UDP |
| Number of packets | ≅1000 |
| Packet length | 512 bytes |
| Flow timeout | 2 s |
| **Mobility Settings** | |
| Mobility of MUs | 300 ms |
| Mobility model of MU | Random way point model |
| Interval time | 0.1 s |
| **Packet Settings** | |
| Packet interval | 100 ms |
| Bit rate | 2 Mbps |
| **Protocol Settings** | |
| Protocol used | IPv6 |
| Latency (processing) | 10 μs |
| **SliceNet Metrics** | |
| Learning rate | 0.001 −0.1 |
| The number of hidden layers | 3 |
| The number of nodes at input layer | 2 |
| The number of nodes at output layer | 2 |
| Activation function | ReLU, and Linear |
| Optimizer | ADAM |
| Number of Epochs | 10 |

The proposed T-S³RA architecture model simulation results are shown in **Fig. 6**. Following the simulation steps, the proposed system with different planes is discussed as shown above.

### B. Comparative Analysis

A comparative analysis section is presented to evaluate the efficiencies of the proposed T-S³RA with respect to the previously investigated methods. A set of significant metrics is considered for comparison with the proposed system. We illustrate the performance of previous network slicing and resource allocation schemes. Existing research works focus on network slicing, resource allocation, or dynamic flow migration (load balancing) during network slicing. Hence, in this study we focused on all three processes along with security because security helps avoids resource wastage in the data and control planes. We compared the performances of the T-S³RA for three slices, eMBB, URLLC, and mMTC, which are represented as S1, S2, and S3, respectively.

*a) Effect on Throughput*

Throughput is an essential metric that is required during communication with the 5G environment. It is defined as the actual rate of information that is being transferred. Furthermore, it is defined as the quantity of data that is transmitted or received per unit of time. In SDN/NFV-based 5G networks, the end user throughput is the number of packets received in bits per second to process a service request.

**Fig. 7** illustrates the performance of throughput with respect to number of slice requests. It is observed that the throughput of the network has higher values in the proposed T-S$^3$RA than in the GRU-DNN [42]. This leads to the involvement of more attackers during network slicing, which increases traffic and delay because each switch and controller component can become congested. SliceNet achieves a better network performance in terms of throughput. Further, throughput increases when using multiple controllers in the global environment. The average throughput of the proposed T-S$^3$RA is 550 Mbps, 467 Mbps, and 437.2 Mbps for S1, S2, and S3, respectively, while the previous GRU-DNN obtained 482.2 Mbps, 495.7 Mbps, and 364.7 Mbps for S1, S2, and S3, respectively.

*b) Effect on Latency*

Latency measures the delay in slice processing. This metric needs to be as small as possible. The QoS of the network can be ensure if low latency is ensured. Because of attackers and compromised nodes in a network, the latency can increase. This is due to the following reasons: (1) Generation of massive traffic at the 5G AP and transmission to the controller and switches, which results in an overflow in the network. This will halt critical services and lead to considerable delay. (2) While the network load increases because of the arrival of more critical service requests, the computation time for each task will increase further. (3) The use of the complex algorithm for traffic scheduling and authentication would induce more computations that will longer time. (4) When slices are selected, the resources may not be allocated, owing to improper management of resource blocks.

As stated above, a comparison of latency performance is depicted in **Fig. 8**. The proposed T-S$^3$RA obtained a lower latency because it uses fast algorithms and effective network slicing and resource allocation. For the DNN, the computation of hyper parameters and tuning requires more time.

*c) Effect on Response Time*

Response time is the time taken to respond to a single slice request. The time required to receive a response from the system should be as low as possible. As shown in **Fig. 9**, minimizing latency helps improve the response time performance. **Fig. 9** also shows the results of the response time comparison. Because T-S$^3$RA is end-to-end secure and effective, we obtained a significantly small response time. The existing work does not use an effective algorithm. The average response times were the lowest in T-S$^3$RA (4.25 s, 2.61 s, and 3.63 s for S1, S2, and S3, respectively). Similarly, for GRU-DNN, the average response times were 6.38 s, 5.4 s, and 5.92 s for S1, S2, and S3, respectively.

*d) Effect on Packet Transmission Ratio*

During the simulations, we observed that the behavior in terms of the ratio of packet transmission requests to the number of devices was good compared to results obtained in the existing work. As the number of devices increases, the ratio of packet transmissions to slice requests increases as well. The traffic is scheduled at the 5G AP using the asymmetric queue model, which increases the packet transmission ratio. In addition, resources are allocated optimally for all slice requests. Thus, we obtained a high packet transmission ratio. The average packet transmission ratios for T-S$^3$RA were 0.89, 0.9945, and 0.854 for S1, S2, and S3, respectively. For GRU-DNN, they were 0.725, 0.78, and 0.696 for S1, S2, and S3, respectively. Among all slices, S2 achieved the highest packet transmission ratio because these services demand highly reliable responses.

*e) Effect on Packet Loss Ratio*

Packet loss is a crucial factor in network performance. For any critical service requests to the network in a global environment, the packet loss must be less than 0.3. **Fig. 11** shows that the packet loss is considerably higher when the switches tend to be overloaded or busy. When the switches are not available, the packets are dropped. This packet loss occurs across network entities on the data plane, edge plane, and the control plane when an excess number of packets arrive at a given time. In an 5G environment with a single controller environment, higher packets losses are often possible.

The packet loss performance is evaluated in terms of the number of devices. The comparison in **Fig. 11** indicates that the packet loss in the proposed T-S$^3$RA is lower than that in GRU-DNN. Because the dynamic flow offloading employed in the data plane balances the arrived flow requests, the packets are not dropped. Similarly, slice requests are handled properly in the network to ensure that sufficient capacity is available for processing the arriving request. Thus, the packet loss ratio is reduced to 0.99 for URLLC, and the maximum packet loss ratio for T-S$^3$RA is 0.25.

*f) Effect on Slice Capacity*

Slice capacity is the load on a particular slice. When the slice value is high, the network obtains a high QoS performance. The main objective of this work is to ensure appropriate slicing for all slice requests that arrive in the network, via identification of the service requirements (throughput, bandwidth, and delay) of the individual slices. **Fig. 12** shows a comparison between slice capacity and number of devices. When we analyze slice capacity, the proposed T-S$^3$RA achieves a high performance. The slice capacity is continuously low for the GRU-CNN.

*g) Effect on Bandwidth Consumption*

Bandwidth is an important constraint for evaluating the performance of slicing under 5G and SDN/NFV-based large-scale networks. In particular, bandwidth is a major resource that is required to perform packet transmission when more packets are received from unauthorized and compromised users. **Fig. 13** shows the bandwidth consumptions of the proposed and existing schemes.

An increase in the number of slice requests leads to high bandwidth consumption to process the requests that have arrived in the network. We demonstrate the performance of the bandwidth consumption with respect to number of slice requests. For GRU-DNN, network slicing and resource allocations are performed without traffic offloading. If a greater number of requests arrive, network slicing and resource allocation can be managed. Here, all users access a single SDN controller for slicing and resource allocation, which therefore consumes larger bandwidth. The proposed scheme uses traffic offloading and a multi-controller environment and, therefore, the bandwidth consumption is comparatively lower; the bandwidth consumption for GRU-DNN is higher owing to the lack of massive traffic handling and single controller problems.

*h) Effect on Slice Acceptance Ratio*

For any slicing, resource allocation, and slice management requests, the slice acceptance ratio is a necessary metric. When the slice acceptance ratio is higher, the user satisfaction rate is higher. In this study, we computed the slice acceptance ratio as a function of the number of slice requests. The processing of each slice and the selection of each slice significantly affects the slice acceptance ratio. **Fig. 14** shows a comparison of the slice acceptance ratios of T-S³RA and GRU-DNN. For T-S³RA, the slice acceptance ratios for all slices consistently increase with respect to the number of slice requests.

A significantly high slice acceptance ratio can be achieved by employing deep-learning-based network slicing and resource allocation. In previous work, the use of deep learning approaches did not result in the best solution. Hence, the slice acceptance ratios were poor. In particular, the problem considers limited parameters for slice selection.

Thus, the proposed T-S³RA achieves better efficiency than the GRU-DNN for all network slices. In addition, the proposed T-S³RA also ensures security while slicing the network and performing resource allocation.

*C. Research Highlights*

In this study, we addressed critical issues encountered in network slicing and resource allocation for devices accessing 5G services. The major highlights are as follows.

A T-S³RA model was designed for network slicing and resource allocation in SDN/NFV-enabled 5G networks. This model is robust, dynamic, reliable, and secure to support mixed types of real-time slices from network equipment.

The T-S³RA model provides the best solution for the deep-learning problem when performing network slicing (SliceNet) and resource allocation (HopFieldNet).

The T-S³RA model allows dynamic flow offloading when the number of slice requests increases. In addition, it does not allow forged slice requests from DDoS attackers.

The T-S³RA model ensures interoperability between the SDN controller, the NFV, and the 5G device while slicing abundant slice requests to appropriate service type networks. The T-S³RA model ensures interoperability as expected in the 5G network because it should be able to support multivendor services or applications. **Table 5** summarizes the algorithm's benefits.

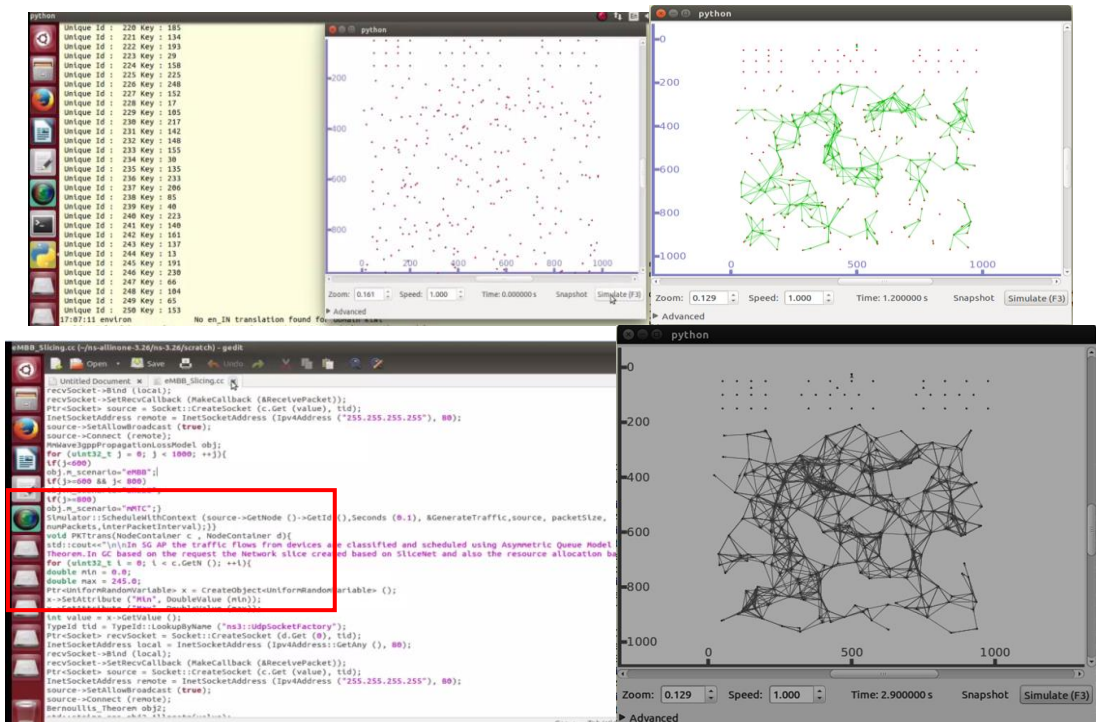

**Fig. 6: Simulation results for node deployment, key generation, and network slicing**

Table 5: Summary of algorithm benefits

| Used operation | Proposed algorithm | Benefits |
|---|---|---|
| Authentication | Boolean logic and PBKDF2 | Enriches the security performance through authentication of devices into the network. It further optimizes the QoS by unaccepting the network slicing and resource allocation. |
| Traffic scheduling | Asymmetric queue model with Bernoulli theorem | Provides crucial features such as Speedup, Scaleup, and Sizeup for any number of slice requests. |
| Network slicing | SliceNet | Provides the appropriate slice for a given user service request, thus maximizing the slice acceptance ratio. It does not introduce computational burden to the SDN controller. |
| Resource allocation | HopFieldNet | Efficient resource allocation maintenance, especially while performing any kind of 5G service (critical and noncritical). Further, it enables increasing of the packet transmission ratio because of sufficient resource allocation. |
| Dynamic offloading | FwBG | Increases the offloading speed and improves scalability. |
| DDoS attackers detection by packet classification | Renyi Entropy | Finds switches that are overloaded and identifies attackers accurately. |

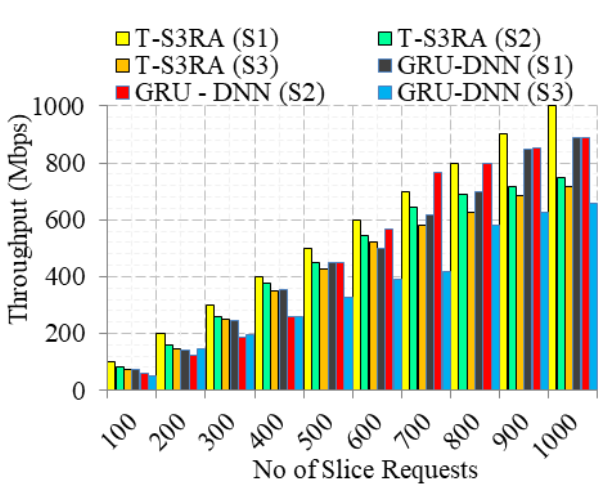

Fig. 7: Throughput vs. number of slice requests

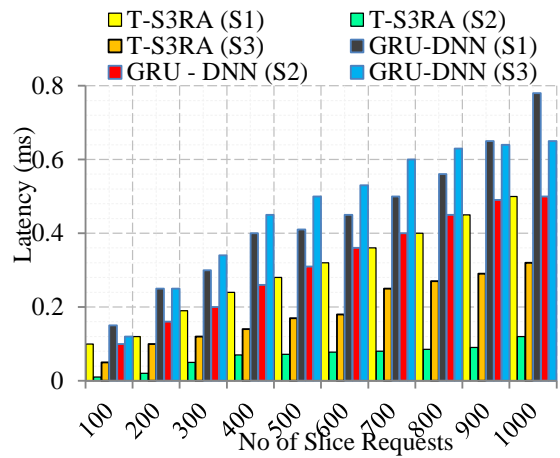

Fig. 8: Latency vs. number of slice requests

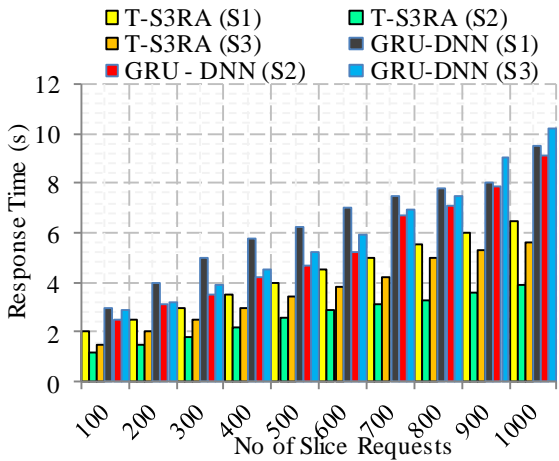

Fig. 9: Response time vs. number of slice requests

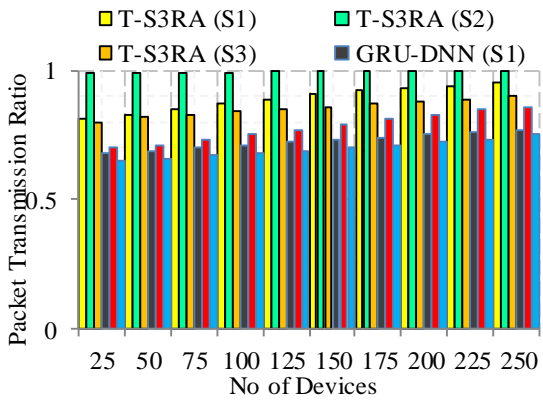

Fig. 10: Packet transmission ratio vs. number of devices

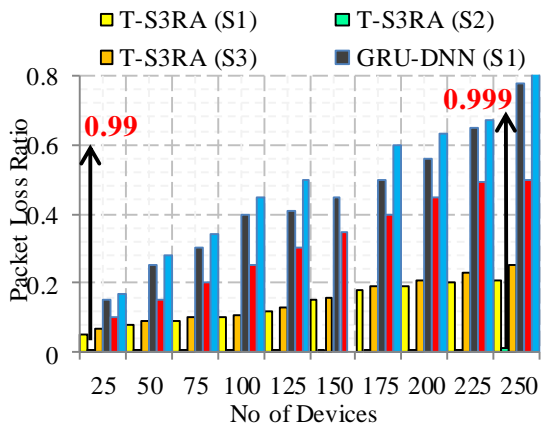

Fig. 11: Packet loss ratio vs. number of devices

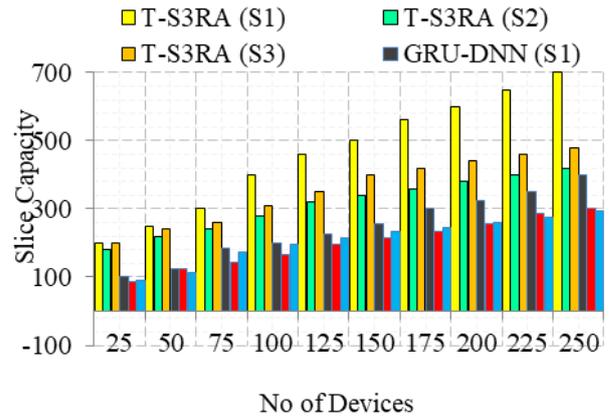

Fig. 12: Slice capacity vs. number of devices

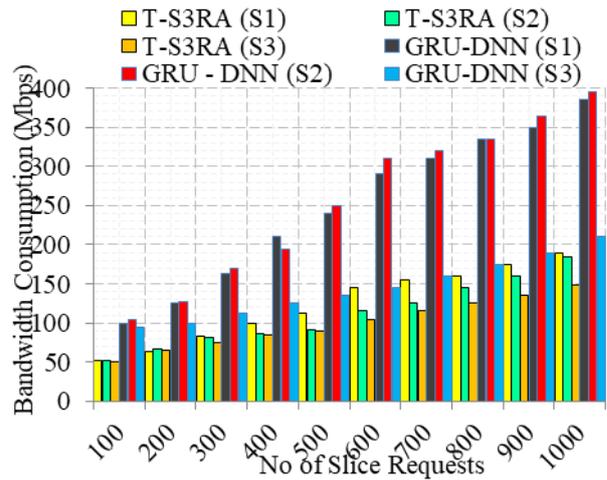

Fig. 13: Bandwidth consumption vs. number of slice requests

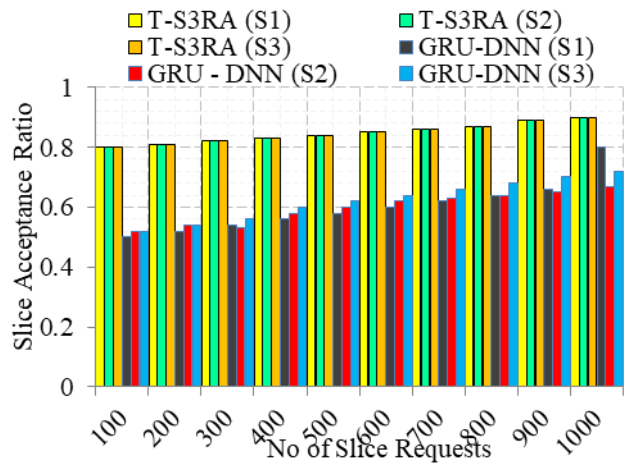

Fig. 14: Slice acceptance ratio vs. number of slice requests

## VI. CONCLUSION

In this study, the QoS was improved in an SDN/NFV-enabled 5G network based on service and SLA requirements for requests arriving from a device or user. The proposed architecture is called T-S$^3$RA and comprises four planes: device, data, local controller, and global controller. The devices or users are authenticated via the VA through 5G AP using PBKDF2. The VA is created and authenticated to the 5G AP for secure communication, and it reduces the communication overhead. Then, the traffic from the 5G AP is classified into two types of queues—HP and LP queues. It is held by the asymmetric queue model using Bernoulli's theorem. The HP request is then forwarded to the LC for network slicing and resource allocation. Here, we proposed SliceNet for performing the slicing; resources are allocated using HopFieldNet. In addition, dynamic flow offloading is implemented using FwBG. The flows are matched with the underloaded switches to avoid packet dropping and to enrich the QoS. Similarly, DDoS attackers are removed from the network via packet classification using Renyi entropy. Finally, the performance of the system was determined for several QoS metrics such as throughput, latency, response time, packet transmission ratio, packet loss ratio, slice capacity, bandwidth consumption, and slice acceptance ratio.